\documentstyle[preprint,aps]{revtex}

\tighten
\input epsf

\begin{document}
\draft

\title{\Large \bf Deterministic Equations of Motion and
Phase Ordering Dynamics}

\author{\bf B. Zheng}

\address{FB Physik, Universit\"at -- Halle, 06099 Halle, Germany}
\address{FB Physik, Universit\"at -- Siegen, 57068 Siegen, Germany}

\maketitle

\begin{abstract}
We numerically solve microscopic deterministic equations of motion for the
2D $\phi^4$ theory with random initial states.
Phase ordering dynamics is investigated.
Dynamic scaling is found and
it is dominated by a fixed point corresponding to the minimum
energy of random initial states. 
\end{abstract}


In recent years
microscopic deterministic equations of motion
(e.g. Newton, Hamiltonian and Heisenberg equations)
have attracted much attention of scientists
in different areas.
 From fundamental view points,
solutions of deterministic equations may describe both
 equilibrium and
non-equilibrium properties of statistical systems,
even though a general proof does not exist,
e.g. see Refs~\cite {fer65,for92,esc94,ant95,cai98}.
Ensemble theories and stochastic equations of motion are 
effective description of static and dynamic properties
of the statistical systems respectively.
With recent development of computers,
it becomes possible 
to solve deterministic equations numerically. 
For example, recently attempt has been made 
for the $O(N)$ vector model
and $XY$ model \cite {cai98,cai98a,leo98}. 
The results support that
 deterministic equations correctly describe
second order phase transitions.
The estimated static critical exponents are consistent with
those calculated from canonical ensembles.
More interestingly, the macroscopic short-time (non-equilibrium)
dynamic behavior of the 2D
$\phi^4$ theory at {\it criticality} has also been investigated 
and dynamic scaling is found
\cite {zhe99,zhe99a}.
The results indicate that deterministic dynamics 
with random initial states
is in a same universality class of 
Monte Carlo dynamics of model A.

On the other hand, 
phase ordering dynamics has been investigated for years
\cite {bra94}.
It concerns how a statistical system evolves into
an ordered phase after a quench from a disordered phase.
For example, the Ising model initially 
at a very high temperature $T_I$
is suddenly quenched
to a temperature $T_F$ well below the critical temperature $T_C$,
and then evolves dynamically.
Because of the competition of the two ordered phases,
it is well known that the equilibration is very slow.
Investigation reveals that
in the {\it late} stage (in microscopic sense) of the dynamic evolution 
there emerges scaling behavior,
which is somehow universal.
The scaling behavior is dominated by
the fixed point $(T_I,T_F)=(\infty,0)$
and away from the fixed point there are corrections to scaling.

Up to now, for simple systems
{\it stochastic} dynamics described by
Langevin-type equations or
Monte Carlo algorithms has been studied.
Scaling behavior of ordering dynamics depends essentially on
whether the order parameter is conserved (model B)
or not (model A).
For the Ising model (or $\phi^4$ theory),
 the dynamic exponent is $z=2$ for model A
and $z=3$ for model B \cite {bra94}.

The purpose of this paper is to study the phase ordering
dynamics with the microscopic 
 deterministic equations of motion,
taking the 2D $\phi^4$ theory as an example.

Following Refs.~\cite {zhe99,cai98}
we consider an isolated system.
The Hamiltonian of the 2D $\phi^4$ theory
on a square lattice is 
\begin{equation}
H=\sum_i \left [ \frac{1}{2} \pi_i^2 
  + \frac{1}{2} \sum_\mu (\phi_{i+\mu}-\phi_i)^2 
  - \frac{1}{2} m^2 \phi_i^2 
  + \frac{1}{4!} g \phi_i^4 \right ]
\label{e10}
\end{equation}
with $\pi_i=\dot \phi_i$ and it leads to the equations of motion
\begin{equation}
 \ddot \phi_i= \sum_\mu (\phi_{i+\mu}+\phi_{i-\mu}- 2\phi_i)
  +  m^2 \phi_i
  - \frac{1}{3!} g \phi_i^3\ .
\label{e20}
\end{equation}
Energy is conserved in these equations. 
Solutions in the long-time regime
are assumed to generate a microcanonical ensemble.
The temperature could be defined as the averaged
kinetic energy. For the dynamic system, however,
the total energy is an even more convenient controlling parameter
of the system, since it is conserved and 
can be input from initial states. 
For given parameters $m^2$ and $g$,
there exists a critical energy density
$\epsilon_c$, separating the ordered phase 
(below $\epsilon_c$) and disordered phase (above $\epsilon_c$).
The phase transition is of the second order.

The order parameter of the $\phi^4$ theory is the magnetization.
The time-dependent magnetization $M\equiv M^{(1)}(t)$  and its second moment
 $M^{(2)}$ are defined as
\begin{equation}
M^{(k)}(t)=\frac {1}{L^{2k}} 
\langle \left [ \sum_i \phi_i(t) \right ]^{(k)} \rangle, \quad k=1, 2.
\label{e30}
\end{equation}
The average is {\it 
over initial configurations} and $L$ is the lattice size. 

Following ordering dynamics with stochastic equations,
we consider a dynamic process that the system initially in
a {\it disordered} state but with energy density
well below $\epsilon_c$ 
is suddenly released to evolve according to Eq.~(\ref {e20}).
For simplicity,
we set initial kinetic energy to 
zero, i.e. $\dot \phi_i(0)=0$.
To generate a random initial configuration $\{\phi_i(0)\}$,
we first fix the magnitude  
$|\phi_i(0)| \equiv c$, then randomly give the sign to $\phi_i(0)$
with the restriction of a fixed magnetization in unit of $c$,
and finally the constant $c$ is determined by
the given energy.
We could also give a distribution
for $|\phi_i(0)|$ but the difference will only be
corrections to scaling.

In case of stochastic dynamics,
 scaling behavior of phase ordering is dominated
by the fixed point $(T_I,T_F)=(\infty,0)$.
In deterministic dynamics,
energy density can not be taken to the real minimum
$e_{min} = - 3 m^4/2 g$  
since the system does not move.
Actually, for the initial states described above,
the energy is given by
\begin{equation}
V=\sum_i \left [ (d - \frac{1}{2} m^2) \phi_i^2 
  + \frac{1}{4!} g \phi_i^4 \right ] \ .
\label{e40}
\end{equation}
Here $d$ is the spatial dimension.
The conjecture is that the scaling behavior
is dominated by the minimum energy density $v_{min}=V_{min}/L^2$,
which is a kind of fixed points.
In this paper, we consider the case of 
$d < m^2/2$. Then
$v_{min} = - 6 (d - m^2/2)^2 /g$.
From now, we redefine the energy density $e_{min}$ 
as zero. Then the fixed point is $\epsilon_0=v_{min}-e_{min}$.

To solve the equations of motion (\ref {e20}) numerically,
 we discretize
$\ddot \phi_i$ by 
$(\phi_i(t+\Delta t)+\phi_i(t-\Delta t)-2\phi_i(t))/(\Delta t)^2$.
After an initial configuration
is prepared,
we update the equations of motion until
$t=650$ or $1000$.
 Then we repeat the procedure with other
initial configurations.
From the experience in Ref.~\cite {zhe99,zhe99a},
$\Delta t=0.05$ is small enough for our
updating times.
In our calculations, we use mainly
a lattice size $L=512$ and samples
of initial configurations
for average are $200$.
Some simulations have also been performed
for $L=1024$ with $50$ samples to estimate the finite size effect.

An important observable is
the equal-time correlation function 
\begin{equation}
C(r,t) =\frac {1}{L^{2}} 
\langle \sum_i \phi_i (t) \phi_{i+r} (t) \rangle \ .
\label{e50}
\end{equation}
Here the lattice site $i+r$ is away from $i$ with a distance $r$.
The scaling hypothesis is that 
at the late stage of the time evolution,
$C(r,t)$ obeys a scaling form
\begin{equation}
C(r,t) = f(r/t^{1/z}) \ ,
\label{e60}
\end{equation}
where $z$ is the so-called dynamic exponent and
the initial magnetization $m_0=0$.
For stochastic dynamics, this scaling form is 
 valid for all temperatures well below
the critical temperature.
Monte Carlo simulations, e.g. for the Ising model,
actually show that at the fixed point
$(T_I,T_F)=(\infty,0)$ the scaling behavior
often emerges at a relatively early time $t$ in the {\it macroscopic}
sense \cite {hum91,fis88}, after a time scale $t_{mic}$
which is large enough in the microscopic sense.
Away from the fixed point, 
there are corrections to scaling.
For deterministic dynamics,
we expect that the minimum energy density of the random initial states
$\epsilon_0=v_{min}-e_{min}$ plays a similar role.

Another interesting observable is the auto-correlation function
\begin{equation}
A(t)=\frac {1}{L^2}\langle \sum_i  \phi_i(0) \phi_i(t) \rangle  .
\label{e70}
\end{equation}
The scaling hypothesis for
 the auto-correlation $A(t)$ 
is a power law behavior 
\begin{equation}
A(t) \sim t^{-\lambda/z}.
\label{e80}
\end{equation}
It implies a divergent correlation time
and ordering dynamics is in some sense {\it `critical'}.
Here $\lambda$ is another independent exponent. 

\begin{figure}[p]\centering
\epsfysize=6.cm
\epsfclipoff
\fboxsep=0pt
\setlength{\unitlength}{0.6cm}
\begin{picture}(9,9)(0,0)
\put(-2,-0.5){{\epsffile{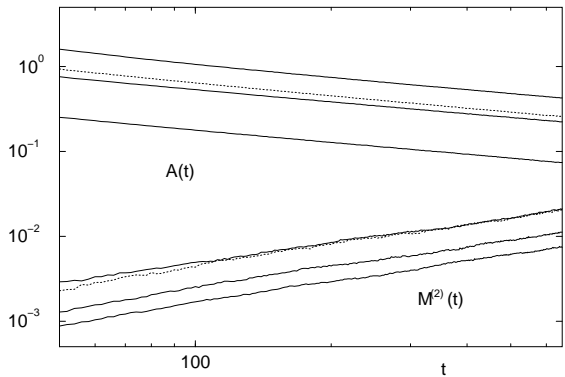}}}
\end{picture}
\caption{The auto-correlation and the second moment with $L=512$
plotted in log-log scale. Solid lines for $A(t)$ are
for $(m^2,g)=(8.0,2.4)$, $(6.0,1.8)$ and $(6.0,5.4)$
(from above) at the fixed points,
while for $M^{(2)}(t)$  are for $(m^2,g)=(6.0,1.8)$, $(8.0,2.4)$ and
$(6.0,5.4)$ (from above). Dashed lines correspond to $(m^2,g)=(6.0,1.8)$
but energy density is $4/3$ above the fixed point.
}
\label{f1}
\end{figure}

We have carried out computations with a lattice size $L=512$
for parameters $(m^2,g)=(6.0,1.8)$, $(6.0,5.4)$ and $(8.0,2.4)$
at the fixed point $\epsilon_0$.
For $(m^2,g)=(6.0,1.8)$, extra simulations
with energy density $\epsilon=\epsilon_0+4/3$
and at $\epsilon_0$ with a large lattice $L=1024$ have
been performed. The auto-correlation has been plotted 
in Fig.~\ref {f1}. The curve for $(m^2,g)=(6.0,1.8)$
with $L=1024$ (not in the figure)  overlaps 
with that for $L=512$. In the figure, we see clearly
a nice power law behavior after 
$t_{mic} \sim 50 - 100$. The dashed line is for $(m^2,g)=(6.0,1.8)$
with energy density $\epsilon=\epsilon_0+4/3$
and correction to scaling is still not so big.
All curves have nearly the same slope indicate
a kind of universality and the fixed point plays an important role.
As is the case of the Ising model with Monte Carlo dynamics
\cite {hum91},
there is a small curvature in the curves, but 
upwards. This gives rise to about one or two percent
 difference of the slope depending on the measured time 
 interval. Slopes for different curves have also
 a comparable uncertainty.
 Taking into account all these factors
 and statistical errors, we estimate
 the exponent $\lambda/z=0.460(10)$. 

\begin{figure}[p]\centering
\epsfysize=6.cm
\epsfclipoff
\fboxsep=0pt
\setlength{\unitlength}{0.6cm}
\begin{picture}(9,9)(0,0)
\put(-2,-0.5){{\epsffile{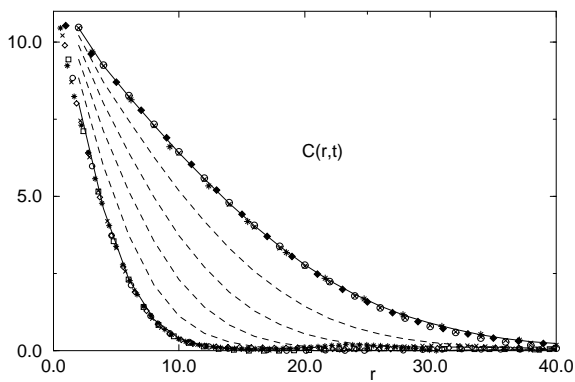}}}
\end{picture}
\caption{Scaling plot for $C(r,t)$.
Six curves for $(m^2,g)=(6.0,1.8)$ at fixed point
with $L=1024$
correspond $t=20$, $40$, $80$, $160$, $320$ and
$640$ (from left).  $\circ$, $\Box$, $\diamond$, $\times$
and $\ast$ fitted to the curve of $t=20$ are those for
$t=40$ to $640$ but $r$ is rescaled according to
$r/t^{1/z}$ with 
$z=2.69$. $\times$ and $\ast$ fitted to the curve of $t=640$ are data
for $(m^2,g)=(6.0,5.4)$
and $(8.0,2.4)$ at the fixed points and $\circ$ 
for $(m^2,g)=(6.0,1.8)$ with energy density $4/3$ 
above the fixed point. The lattice size is $L=512$ and
both axes are rescaled with suitable constants.
Full diamonds represent the scaling function for
the Ising model at the zero temperature.
}
\label{f2}
\end{figure}

In Fig.~\ref {f2}, the equal-time correlation function $C(r,t)$ 
is displayed. The curves are for $(m^2,g)=(6.0,1.8)$
with $L=1024$ and one sees clear self-similarity 
during time evolution. 
According to the scaling form (\ref {e60}), 
data for different time t should collapse if
$r$ is suitably rescaled by $t^{1/z}$.
In other words, searching for the best collapse
of the data we can obtain the dynamic exponent $z$. 
This collapse of the data is shown 
on the first curve from the left.
All data points locate nicely on a curve
except for a small departure for $t=20$.
The corresponding dynamic exponent measured from
a time interval $[40,640]$ is $z=2.69(9)$.
In Table~\ref {t1}, we list values of $z$ for different parameters and
measured in different time intervals. 
Again, for larger time $t$
the dynamic exponent $z$ tends to be slightly smaller.
We believe the small deviation for different
parameters $(m^2,g)$ is more or less due to uncontrolled systematic
errors or/and possible corrections to scaling.
From the table, we estimate the dynamic exponent $z=2.65(10)$.
This is significantly different
 from $z=2.0$ for the Ising model
with stochastic dynamics of model A.

Very interesting is that the scaling function $f(x)$
in Eq.~(\ref {e60}) for the $\phi^4$ theory
 is the same as that of the Ising model 
with Monte Carlo dynamics of model A
\cite {hum91,maz90}, even though
the exponent $z$ is different. This is shown 
on the last curve from left in Fig.~\ref {f2}.
To plot the functions, $r$ and $C(r,t)$ have been suitably rescaled
by constants. We did not try
to get a `best' fit to all the data but only to show
they are indeed a same function.
For the data of $(m^2,g)=(6.0,5.4)$ ($\times$) 
and $(6.0,5.4)$ ($\circ$),
only $r$ is rescaled. 
For the Ising model (full diamonds),
the rescaling factor for $r$ happens to be $1/2$.

A simple understanding of the scaling behavior of
$C(r,t)$ can be achieved from the second moment
of the magnetization.
Integrating
over $r$ in Eq.~(\ref {e60}), we obtain a power law behavior 
\begin{equation}
M^{(2)}(t) \sim t^{d/z} \ .
\label{e90}
\end{equation}
This is shown also in Fig.~\ref {f1}.
Even though there are some visible fluctuations,
power law behavior is observed.
From slopes of the curves after $t \sim 100$, we measure
the exponent $d/z=0.76(3)$. Then we estimate the dynamic
exponent $z=2.63(10)$.

For discussions above, the initial magnetization $m_0$
is zero. If $m_0$ is a non-zero,
the system reaches a unique ordered state within a finite time.
If $m_0$ is infinitesimal small, however,
the time for reaching the ordered state is also infinite
and scaling behavior can still be expected, at least 
at relatively early times (in macroscopic sense).
In this case, an interesting observable is the magnetization itself
and at early times it increases by a power law
\begin{equation}
M(t) \sim t^{\theta}, \quad \theta=(d-\lambda)/z.
\label{e100}
\end{equation}
The exponent $\theta$ can be written as $x_0/z$, with $x_0$
being the scaling dimension of $m_0$.
This power law behavior has deeply been investigated
in critical dynamics \cite {jan89,zhe98}.

\begin{figure}[p]\centering
\epsfysize=6.cm
\epsfclipoff
\fboxsep=0pt
\setlength{\unitlength}{0.6cm}
\begin{picture}(9,9)(0,0)
\put(-2,-0.5){{\epsffile{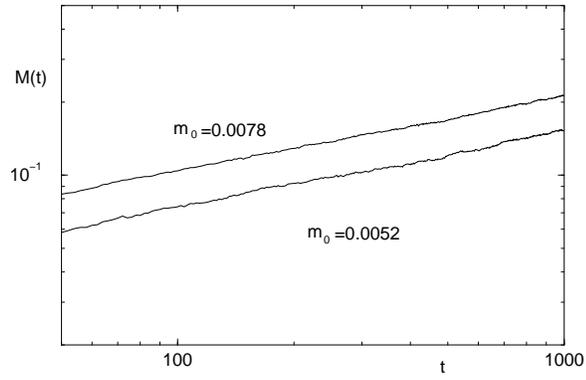}}}
\end{picture}
\caption{The magnetization in log-log scale.
The lattice size is $L=512$.
}
\label{f3}
\end{figure}

In Fig.~\ref {f3}, the initial increase of the magnetization
is shown. After $t_{mic} \sim 80$, nice power law behavior is seen.
To avoid finite $m_0$ effect, very small values of
$m_0$ have been chosen. The resulting exponent $\theta$
is $0.308(9)$ and $0.315(30)$ for $m_0=0.0078$ and
$0.0052$ respectively. Taking into account the errors,
we consider $\theta=0.308(9)$ as the final result.
With $\theta$ and $\lambda/z$ in hand,
from the scaling relation $\theta=(d-\lambda)/z$
again we can calculate the dynamic exponent $z=2.60(5)$.

In Table~\ref {t2}, we have summarized
all the measurements of the exponents. 
The agreement of  
different measurements of $z$ strongly supports
the dynamic scaling hypothesis.
The exponents of the Ising model with stochastic
dynamics of model A are from theoretical calculations
\cite {fis88,bra94}.
In Monte Carlo simulations,
there may be small deviation
\cite {fis88,fur90,hum91}.
It is interesting that the dynamic exponent $z$ for the $\phi^4$ 
theory with deterministic dynamics
is clearly different from
that of the Ising model with stochastic
dynamics but the exponent $\lambda$
looks the same.

In Refs.~\cite {zhe99,zhe99a,zhe98}, we know that 
in dynamic critical phenomena,
deterministic dynamics 
for the 2D $\phi^4$ theory
and stochastic dynamics of model A
for the Ising model 
are in a same universality class.
Why is it not the case in ordering dynamics?
This may be traced back to the energy conservation
in our deterministic equations.
Since energy couples to the order parameter,
deterministic dynamics is somehow believed to be a realization
of model C \cite {hoh77}. For critical dynamics, in two-dimensions
model A and model C are the same.
For ordering dynamics, however, 
model A and model C can be different.
It is pointed out in Ref.~\cite {hoh77} that
in many cases real physical systems may be intermediate
between model A and C.

When $d-m^2/2$ becomes positive,
 $v_{min}$ moves to zero. This is an unnormal fixed point
 ($\phi_i(0) \equiv 0$), from which
the system can not move.
Around this fixed point, self-similarity is also 
observed in time evolution, but a simple scaling form
as Eq.~(\ref {e60}) does not give good collapse of the data,
at least up to the time $t=650$.
Further understanding remains open. 

In conclusions, we have investigated 
ordering dynamics
governed by deterministic equations of motion,
taking the 2D $\phi^4$ theory as an example.
Scaling behavior is found and it is dominated by
the fixed point corresponding to the minimum energy
of random initial states.
The dynamic exponent $z$ is 
different from that of stochastic dynamics of model A,
while the scaling function for the equal-time correlation
$C(r,t)$ is the same.  
Deterministic dynamics with energy conservation
might be a realization of model C.

{\bf Acknowledgements}:
Work supported in part by the Deutsche Forschungsgemeinschaft
under the project TR 300/3-1. 


\begin{thebibliography}{10}

\bibitem{fer65}
{E. Fermi, J. Pasta and S. Ulam},  in {\em {Collected Papers of Enrico Fermi}},
  edited by {E. Segr\'e} (Univ. Chicago, Chicago, 1965).

\bibitem{for92}
{J. Ford}, Phys. Rep. {\bf {213}},  271  (1992).

\bibitem{esc94}
{D. Escande, H. Kantz, R. Livi and S. Ruffo}, J. Statist. Phys. {\bf {76}},
  605  (1994).

\bibitem{ant95}
{M. Antoni and S. Ruffo}, Phys. Rev. {\bf {E52}},  2361  (1995).

\bibitem{cai98}
{L. Caiani, L. Casetti and M. Pettini}, J. Phys. {\bf {A31}},  3357  (1998).

\bibitem{cai98a}
{L. Caiani, L. Casetti, C. Clementi, G. Pettini, M. Pettini and R. Gatto},
  Phys. Rev. {\bf {E57}},  3886  (1998).

\bibitem{leo98}
{X. Leoncini and A.D. Verga}, Phys. Rev. {\bf {E57}},  6377  (1998).

\bibitem{zhe99}
{B. Zheng, M. Schulz and S. Trimper}, Phys. Rev. Lett. {\bf {82}},  1891
  (1999).

\bibitem{zhe99a}
{B. Zheng}, {\em Microscopic Deterministic Dynamics and Persistence Exponent},
  {Univ. Halle}, 1999, to be published in Mod. Phys. Lett. B.

\bibitem{bra94}
A.~J. Bray, Adv. Phys. {\bf {43}},  357  (1994), and references therein.

\bibitem{hum91}
K. Humayun and A.~J. Bray, J. Phys. {\bf {A24}},  1915  (1991).

\bibitem{fis88}
{D. S. Fisher and D. A. Huse}, Phys. Rev. {\bf {B38}},  373  (1988).

\bibitem{maz90}
{G. F. Mazenko}, Phys. Rev. {\bf {B42}},  4487  (1990).

\bibitem{jan89}
{H. K. Janssen, B. Schaub and B. Schmittmann}, Z. Phys. {\bf {B 73}},  539
  (1989).

\bibitem{zhe98}
B. Zheng, Int. J. Mod. Phys. {\bf B12},  1419  (1998), review article.

\bibitem{fur90}
{H. Furukawa}, Phys. Rev. {\bf {B42}},  6438  (1990).

\bibitem{hoh77}
{P.C. Hohenberg and B.I. Halperin}, Rev. Mod. Phys. {\bf {49}},  435  (1977).

\end{thebibliography}

\begin{table}[h]\centering
\begin{tabular}{cccccc}
  $(m^2,g)$ &  & $t_1=40$ &   $t_1=80$ &  $t_1=160$ &  $t_1=320$     \\        
  (8.0,2.4) &        & 2.84(5)  &   2.83(6)  &  2.76(10)  &  2.79(12)\\
  (6.0,5.4) &        & 2.75(2)  &   2.75(1)  &  2.70(1)   &  2.72(1)\\
  (6.0,1.8) &        & 2.78(3)  &   2.76(1)  &  2.67(3)   &  2.57(4)\\
            & L=1024 & 2.72(7)  &   2.69(9)  &  2.67(7)   &  2.66(4) \\  
            & $\epsilon=\epsilon_{0}+4/3$    
                     & 2.78(5)  &   2.74(8)  &  2.70(10)  &  2.69(10)\\
\end{tabular}
\caption{The dynamic exponent $z$ estimated from scaling collapse
of $C(r,t)$ in a time interval $[t_1,640]$. If not specified,
the lattice size $L=512$ and the energy density is at its
fixed points.
}
\label{t1}
\end{table}

\begin{table}[h]\centering
\begin{tabular}{c|cc|ccc|c}
         &          &                      &                  &  $z$ 
                    &  &\\
\hline
         & $\theta$ & $\lambda/z$          & $d/(\lambda/z+\theta)$ & $C(r,t)$
                 & $M^{(2)}$     & $\lambda$       \\
\hline
$\phi^4$ & 0.308(9)    &   0.460(10)   &    2.60(5)    &  2.65(10)
                 &   2.62(10)    &  1.22(5)     \\
\hline
Ising  &         & 0.625           &                &    2
                 &               &  1.25  \\
\end{tabular}
\caption{Exponents of the $\phi^4$ with deterministic dynamics.
To calculate $\lambda$,
$z$ measured from $C(r,t)$ is taken as input.
Values for the Ising model are theoretical results with
stochastic dynamics of model A
\protect\cite {fis88,bra94}. 
}
\label{t2}
\end{table}

\end{document}